\documentclass[twocolumn]{aastex701}

\usepackage{booktabs}
\usepackage{tcolorbox}

\shorttitle{Semantic Search for Galaxy Images}
\shortauthors{Koblischke et al.}

\begin{document}

\title{Semantic Search for 100M+ Galaxy Images Using AI-Generated Captions}

\correspondingauthor{Nolan Koblischke}

\author[0000-0001-5396-5824]{Nolan Koblischke}
\altaffiliation{Work performed in part at New York University.}
\affiliation{David A. Dunlap Department of Astronomy and Astrophysics, University of Toronto, 50 St. George Street, Toronto, ON, M5S 3H4, Canada}
\affiliation{Dunlap Institute for Astronomy \& Astrophysics, University of Toronto, 50 St. George Street, Toronto, ON, M5S 3H4, Canada}
\email[show]{nolan.koblischke@mail.utoronto.ca}

\author[0009-0007-4952-1674]{Liam Parker}
\affiliation{Berkeley Center for Cosmological Physics, Department of Physics, University of California, Berkeley, CA 94720, USA}
\affiliation{Center for Data Science, New York University, 60 5th Ave, New York, NY 10011, USA}
\affiliation{Lawrence Berkeley National Lab, 1 Cyclotron Road, Berkeley, CA 94720, USA}
\affiliation{Center for Computational Astrophysics, Flatiron Institute, 162 Fifth Avenue, New York, NY 10010, USA}
\email{lhparker@berkeley.edu}

\author[0000-0001-7956-0542]{Francois Lanusse}
\affiliation{Universit\'e Paris-Saclay, Universit\'e Paris Cit\'e, CEA, CNRS, AIM, F-91191 Gif-sur-Yvette, France}
\affiliation{Center for Computational Astrophysics, Flatiron Institute, 162 Fifth Avenue, New York, NY 10010, USA}
\email{francois.lanusse@cnrs.fr}

\author[0000-0001-6855-442X]{Jo Bovy}
\affiliation{David A. Dunlap Department of Astronomy and Astrophysics, University of Toronto, 50 St. George Street, Toronto, ON, M5S 3H4, Canada}
\affiliation{Dunlap Institute for Astronomy \& Astrophysics, University of Toronto, 50 St. George Street, Toronto, ON, M5S 3H4, Canada}
\email{jo.bovy@utoronto.ca}

\author[0000-0002-1790-7536]{Irina Espejo}
\affiliation{Center for Data Science, New York University, 60 5th Ave, New York, NY 10011, USA}
\email{iespejomorales@simonsfoundation.org}

\author[0000-0002-1068-160X]{Shirley Ho}
\affiliation{Center for Computational Astrophysics, Flatiron Institute, 162 Fifth Avenue, New York, NY 10010, USA}
\affiliation{Department of Physics \& Center for Data Science, New York University, 60 5th Ave, New York, NY 10011, USA}
\affiliation{Department of Astrophysical Sciences, Princeton University, 4 Ivy Lane, Princeton, NJ 08544, USA}
\email{shirleyho@flatironinstitute.org}

\begin{abstract}
Finding scientifically interesting phenomena through slow manual labeling campaigns severely limits our ability to explore the billions of galaxy images produced by telescopes. In this work, we develop a pipeline to create a \textit{semantic} search engine from completely unlabeled image data. Our method leverages Vision-Language Models (VLMs) to generate descriptions for galaxy images, then contrastively aligns a pre-trained astronomy foundation model with these embedded descriptions to produce searchable embeddings at scale. We find that current VLMs provide descriptions that are sufficiently informative to train a semantic search model that outperforms direct image similarity search. Our model, AION-Search, achieves state-of-the-art zero-shot performance on finding rare phenomena despite training on randomly selected images with no deliberate curation for rare cases. Furthermore, we introduce a VLM-based re-ranking method that nearly doubles the recall for our most challenging targets in the top-100 results. For the first time, AION-Search enables flexible semantic search for over 100 million galaxy images, enabling discovery from previously infeasible searches, including the identification of 36 new extragalactic stellar stream candidates. More broadly, our work provides an approach for making large, unlabeled scientific image archives semantically searchable, expanding data exploration capabilities in fields from Earth observation to microscopy. The code, data, and app are publicly available at \url{https://github.com/NolanKoblischke/AION-Search}.
\end{abstract}


\section{Introduction}
Recent advances in AI capabilities have prompted visions of ``a country of geniuses in a datacenter'', highly capable AI agents available on demand to accelerate scientific discovery~\citep{amodei2024machines}. In astrophysics, where telescopes will generate billions of galaxy images that far exceed human capacity for manual inspection, deploying such AI agents at scale presents an opportunity to analyze vast astronomical datasets. Here, we take a step toward this vision by using Vision-Language Models (VLMs) to describe galaxy images in natural language.

VLMs, including systems such as GPT~\citep{gpt4osystemcard} and Gemini~\citep{comanici2025gemini25pushingfrontier}, are neural networks trained to connect visual inputs with language, enabling tasks such as image captioning and visual question answering. At a high level, many modern VLMs are built by coupling a vision encoder to a language model, then training the system on large collections of paired images and text. In this work, we use VLMs as scalable annotators to generate descriptions of galaxy images. These descriptions serve as training data for AION-Search, a CLIP-based~\citep{clip} semantic search engine.

PAPERCLIP~\citep{mishra-sharma2024paperclip} pioneered astronomical text-image retrieval using $\sim$4,000 abstracts and associated observations for Hubble Space Telescope observing proposals. However, their approach was limited to a corpus of abstracts and misaligned text-image pairs, where one text can map to multiple images or contain irrelevant information. With VLM annotators, we can overcome these constraints by generating specific descriptions for each image, enabling semantic search for any telescope at scale. As wide-field imaging surveys collectively image hundreds of millions of galaxies, finding specific phenomena of interest remains challenging due to the sheer data volume. This challenge will only grow with upcoming surveys such as Euclid~\citep{euclidredbook} and LSST~\citep{lsst}. We demonstrate our approach on the Legacy Survey~\citep{legacy} and Hyper Suprime-Cam (HSC;~\citealt{hsc}), but the method generalizes to any imaging survey.

Current approaches to finding specific galaxy types rely on citizen-science labeling campaigns~\citep{walmsley_tidal,gonzalez2025discoveringstronggravitationallenses}. For Galaxy Zoo, volunteers answer hierarchical fixed-choice questions such as ``Is there any sign of a spiral arm pattern?'' followed by ``How many spiral arms are there?''~\citep{2022walmsley,euclidcollaboration2025euclidquickdatarelease,lintott_galaxyzoo}. With upcoming surveys producing billions more galaxy images, these manual approaches will inevitably miss discoveries. These labeling campaigns have been used to train supervised deep-learning models~\citep{dieleman2015,walmsley2023gzdesi,2022walmsley}. However, such models inherit the biases of the volunteer-labeled training set and are constrained to predicting answers to fixed question sets, and therefore do not support free-form, open-vocabulary queries.

\begin{figure*}[t]
    \centering
    \includegraphics[width=\linewidth]{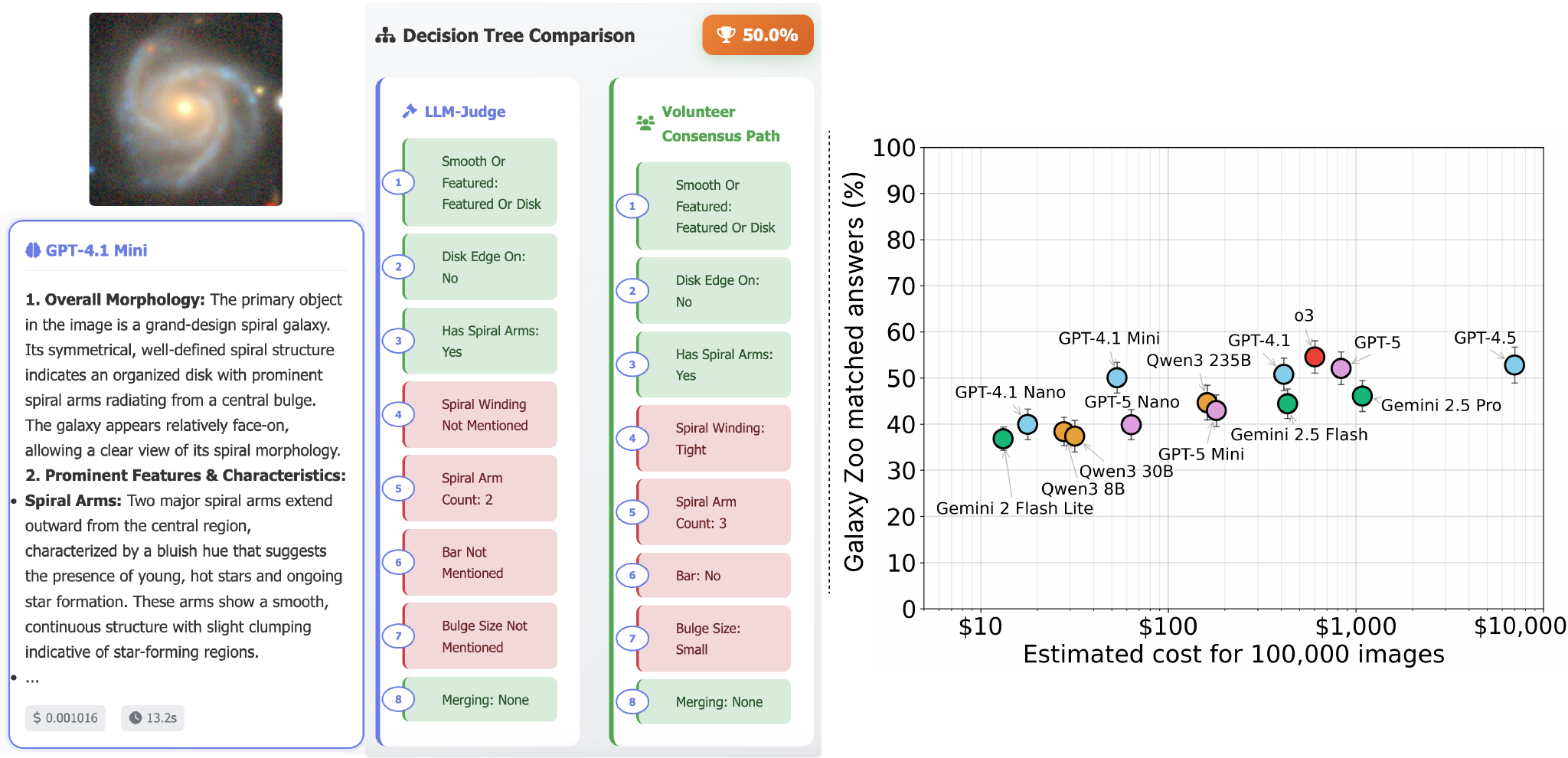}
    \caption{Evaluating VLM descriptions against Galaxy Zoo consensus labels. \emph{Left:} GPT-4.1-mini generates a free-form description for a galaxy image. An LLM judge converts the description into answers to the Galaxy Zoo decision-tree questions, which are then compared to the volunteer-consensus answers. \emph{Right:} For each model, we plot the Galaxy Zoo answer match rate against the estimated batch-API cost to caption 100,000 images. Answer match rate is the fraction of individual decision-tree answers inferred from the free-form descriptions that match the volunteer-consensus answers. Points show the mean over three runs (GPT-4.5: one run) with standard errors. GPT-4.1-mini provides the strongest accuracy-cost trade-off within our captioning budget.}
    \label{fig:vlm_description_bench}
\end{figure*}
While self-supervised methods offer similarity search~\citep{stein,AstroCLip,hayat2021}, they retrieve the most visually similar images to the query image, rather than all images containing a desired feature, rendering them insufficient for targeted searches. An effective semantic search engine would require informative captions that capture the physical phenomena present in galaxy images. VLMs demonstrate promising capabilities for this task, having been pretrained on billions of image-text pairs scraped from the web~\citep{schuhmann2022laion5bopenlargescaledataset}, which likely include research figures, observational annotations, and public discourse about galaxy images. This allows them to describe a broad range of observable phenomena rather than the predefined categories derived from volunteer labeling campaigns typically used for supervised training. Our objective is to evaluate whether VLM-generated descriptions possess sufficient accuracy and information content to serve as training data for a contrastive learning framework. To enable efficient semantic search at scale, we train a CLIP-based model that aligns image representations from AION~\citep{2025aion1}, an astronomy foundation model, with VLM-generated descriptions. With the flexibility provided by AION and VLM descriptions, this approach provides scalable semantic search extensible to future telescopes.
\subsection{Related Work}
In recent years, approaches to improve text-visual understanding, such as CLIP~\citep{clip}, have achieved remarkable performance on a variety of downstream tasks, including retrieval, by training on pairs of image-captions extracted from the web. However, because high-quality image-text pairs are finite, recent work has explored using synthetic captions to augment training data~\citep{zheng2024dreamliplanguageimagepretraininglong, zhang2025syntheticcaptions}. For CLIP-like models, SynthCLIP~\citep{hammoud2024synthclipreadyfullysynthetic} demonstrates that while performance with fully synthetic captions does not match performance with real captions, it exhibits good scalability trends and has competitive performance in some of the downstream tasks. For retrieval, CLIPS~\citep{liu2024clipsenhancedclipframework} proposes to learn both with web and synthetically augmented captions, achieving state-of-the-art in cross-modal retrieval tasks. In astrophysics, \citet{vago2026augmentingrepresentationsscientificpapers} uses a CLIP framework to align X-ray spectra with summaries from scientific literature, demonstrating cross-modal retrieval.

In the context of scientific images, the scarcity of text-image pairs is even more pronounced because it relies on domain expertise and standardized benchmarks are rare. A common workaround is to convert categorical labels into natural-language prompts with templates and train CLIP-style models on these synthetic pairs~\citep{liu2024remoteclipvisionlanguagefoundation,Zhenwei,khattak2024unimedclipunifiedimagetextpretraining}. In particular, INQUIRE~\citep{inquire} produces a benchmark for text-to-image retrieval in ecology and additionally evaluates re-ranking with multimodal models. More recently, a few efforts experiment with generating captions directly from images using pre-trained VLMs~\citep{chen2025lrsclipvisionlanguagefoundationmodel,he2025enhancingremotesensingvisionlanguage,YuanChatEarthNet}. VLMs have also been used in astronomy for image classification~\citep{Tanoglidis_2024,zaman2025astrollavaunificationastronomicaldata}. To our knowledge, this work is the first example in astrophysics of using VLM-generated captions to generate training sets for semantic retrieval.
\subsection{Paper Outline}
This paper is organized as follows. In \S\ref{sec:methodology}, we benchmark VLM descriptions against human annotations, generate captions at scale, and train a contrastive model to produce searchable image embeddings. In \S\ref{sec:results}, we evaluate retrieval performance for spiral galaxies, mergers, and gravitational lenses, test a VLM re-ranking stage, compare against supervised classifiers, and apply the system to discover new extragalactic stellar streams. We discuss implications and limitations in \S\ref{sec:discussion}.

\begin{figure*}[t]
  \centering
  \begin{tcolorbox}[colback=gray!10, colframe=black, title=Training set generation prompt]
Analyze this astronomical image and provide a detailed description, assuming the reader has domain expertise and the description will be reviewed by astrophysicists.

\vspace{4pt}
1. Morphology: Describe a detailed morphological classification of the main object(s) (e.g., spiral, elliptical, irregular, merging system, etc.).
\vspace{4pt}

2. Prominent Features \& Characteristics: Detail specific astronomical features observed, for example: spiral arms, bars, dust lanes, star-forming regions, tidal features, merger remnants, foreground stars, lensing effects, etc. Mention these features if and only if they are present in the image. If they are not present, do not mention them. The features mentioned here were just examples, state any feature you believe astrophysicists would be interested in. For each feature, describe its key characteristics (e.g., morphology, extent, brightness, orientation, interaction with other components, physical phenomena occurring, etc.).
\vspace{4pt}

3. Additional Context \& Interpretation: Include any further observational details or astrophysical interpretations.
\vspace{4pt}

Only focus on the center object(s) in the image. Keep your response under 300 words.
\end{tcolorbox}
\caption{Prompt used to generate the VLM descriptions that form the training set for AION-Search. This prompt was optimized using our Galaxy Zoo benchmark (\S\ref{sec:gz_curation}) while remaining general to capture features beyond those in Galaxy Zoo questionnaires.}
\label{fig:prompt}
\end{figure*}

\section{Methodology}\label{sec:methodology}
We investigate the visual understanding capabilities of VLMs in astronomy, whose pretraining corpora include astronomical literature and imagery at web scale~\citep{schuhmann2022laion5bopenlargescaledataset,olmo2025olmo3}. We first quantify this with a benchmark against human annotations (\S\ref{sec:gz_curation}), then use the best-performing VLM to generate galaxy image descriptions that serve as training data for a contrastive model producing searchable embeddings directly from images.
\subsection{Evaluating vision-language model image descriptions}\label{sec:gz_curation}
We evaluate the capacity of Vision-Language Models to generate scientifically accurate descriptions of galaxy images using the Galaxy Zoo-DECaLS catalog~\citep{2022walmsley}, which provides crowd-sourced annotations for Legacy Survey images, where human volunteers answer a series of hierarchical questions about each galaxy's visual properties. We use this dataset as a testbed to evaluate relative performance across models in capturing key features of galaxies, rather than as a comprehensive benchmark. For cost-effective evaluation, we curate a test set of 64 images meeting the criteria: (i) each question having strong consensus ($>$70\% agreement) among sufficient annotators ($>$10), and (ii) diversity, enforced by limiting any identical classification to at most five examples. This size enables multiple evaluation runs across 14 different VLMs while keeping API costs manageable (with the most expensive model, GPT-4.5, costing \$9 per evaluation when this evaluation was run) and provides sufficient statistical power to distinguish broad performance tiers and identify cost-effective models, though fine-grained differences between adjacent models may not be significant.

\begin{figure}[t]
    \centering
    \includegraphics[width=\linewidth]{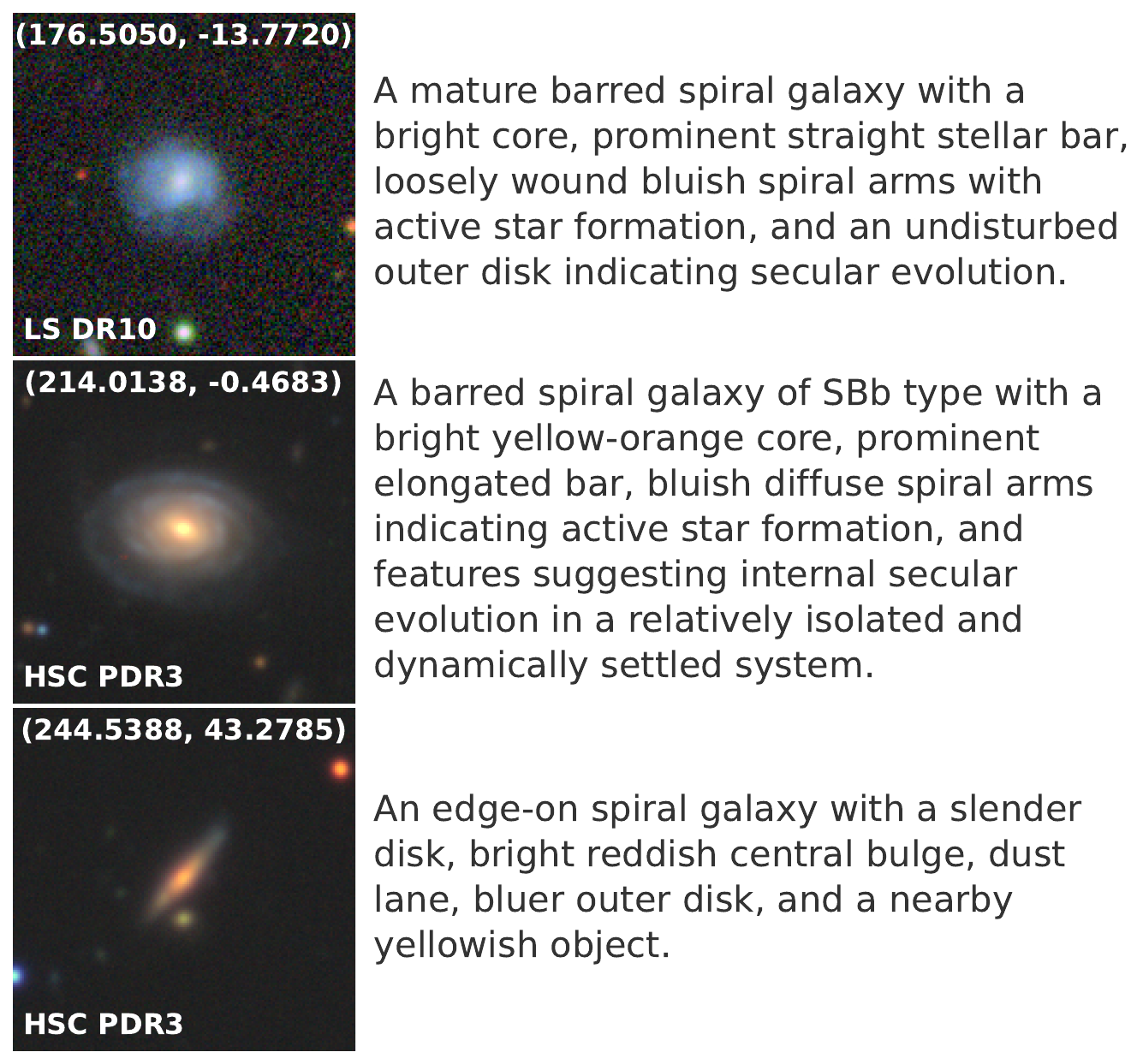}
    \caption{Example galaxy images and their GPT-4.1-mini descriptions. The generated captions generally capture the broad morphology visible in each image, though fine-grained details can be incorrect. The top image was chosen as a failure case: the model describes a bar and loosely wound arms that are not apparent. The bottom two are randomly selected. Caption accuracy is evaluated quantitatively in \S\ref{sec:gz_results}.}
    \label{fig:example_galaxies}
\end{figure}

We conduct two forms of evaluation. First, we directly prompt VLMs to answer the same questions that human annotators answered. Second, we prompt VLMs to generate free-form image descriptions, since the same prompt will generate training data for our contrastive model, which we subsequently evaluate using an LLM-judge to extract answers to the annotation questions. Specifically, we use Gemini-2.5-Flash~\citep{comanici2025gemini25pushingfrontier} to extract decision tree answers from the generated descriptions, allowing responses of ``not stated'' when information is absent, which are scored as incorrect. Figure~\ref{fig:vlm_description_bench} shows the results of this evaluation using the prompt in Figure~\ref{fig:prompt}, demonstrating that accuracy generally trades off with cost. We use this evaluation to select our captioning model, GPT-4.1-mini, which offers the strongest Galaxy Zoo answer match rate ($50.1\pm3.3$\%) within our cost budget. Figure~\ref{fig:example_galaxies} shows example images and their generated descriptions. VLM description accuracy results are discussed further in \S\ref{sec:gz_results}.

\begin{figure*}[t]
    \centering
    \includegraphics[width=0.7\linewidth]{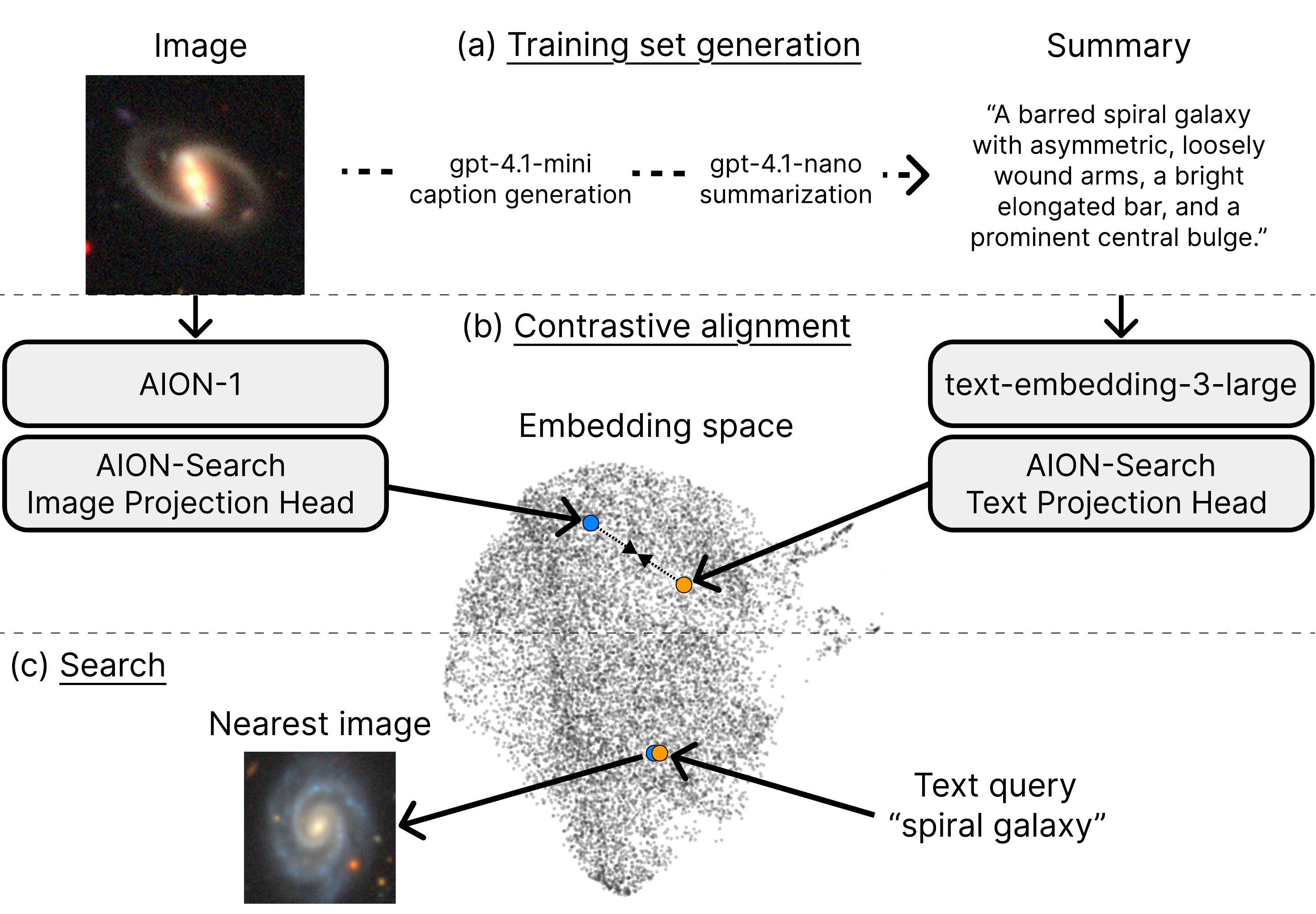}
    \caption{Overview of the AION-Search pipeline. (a) Each image is captioned by GPT-4.1-mini using the prompt in Figure~\ref{fig:prompt}, and GPT-4.1-nano summarizes the caption into a single sentence. (b) To create a shared image--text embedding space, images are embedded with AION-1~\citep{2025aion1} and text summaries are embedded with \texttt{text-embedding-3-large}~\citep{openai_text_embedding_3_large_announcement}. These embeddings are passed through multilayer perceptron (MLP) projection heads trained with a contrastive loss, which pulls matching image--text pairs together and pushes mismatched pairs apart. (c) At search time, a text query is embedded and passed through the learned text projection head. The retrieved galaxies are those whose projected image embeddings have the highest cosine similarity to the query embedding.}
    \label{fig:overview}
\end{figure*}

\subsection{Caption generation for galaxy images}\label{sec:caption_generation}
To construct our training dataset, we sample 300,000 galaxy images brighter than 19 magnitudes in $r$-band. We keep the selection simple, a single brightness cut to remove faint sources, and we do not build the training set from curated rare-object catalogs (e.g., Galaxy Zoo or strong lens lists), so the model stays general rather than biased toward any one target class. We split our sampling across telescopes to demonstrate that our search can work across different imaging modalities, 120,000 from HSC (PDR3 Wide; \citealt{Aihara_2022}) and 180,000 from Legacy Survey (DR10 South; \citealt{legacy}). These images are drawn from the Multimodal Universe (MMU) dataset, which has been preprocessed to remove non-galaxy sources and low-quality images~\citep{MultimodalUniverse2024}. The final dataset consists of 160$\times$160 pixel cutouts with five channels for HSC and four channels for Legacy Survey. We turn each multi-band cutout into a color image for the VLMs by rescaling the bands, mapping ($z$,$r$,$g$) bands to (R,G,B), and applying an arcsinh stretch~\citep{Lupton_2004} to enhance faint features. Using the optimal prompt–model configuration identified through our evaluation framework (i.e., GPT-4.1-mini with the prompt in Fig.~\ref{fig:prompt}), we generate descriptions for each image. These textual descriptions are subsequently embedded using \texttt{text-embedding-3-large}~\citep{openai_text_embedding_3_large_announcement}. After excluding overlap with downstream benchmarks, 255,948 galaxies receive embedded descriptions and are used for training, with generation and embedding costs totaling $\sim$\$150 through the OpenAI API.

\subsection{Learning semantic embeddings through contrastive alignment}\label{sec:training_clip}
These generated embeddings already enable semantic search. However, writing descriptions for all images in the Legacy Survey and HSC MMU catalogs would be prohibitively expensive. To address this, we predict text-searchable embeddings directly from images using the approach summarized in Figure~\ref{fig:overview}. We use AION~\citep{2025aion1} as our image encoder, an astronomy foundation model pretrained via masked modeling on over 200 million observations that achieves state-of-the-art performance on galaxy image similarity search. The AION Transformer-based encoder-decoder architecture accommodates varying input formats, enabling it to handle both HSC images (five bands) and Legacy Survey images (four bands), and can be used for future telescope surveys. We use AION-1-Base, the 300M parameter variant with 768-dimensional embeddings, contrastively aligning its image encoder outputs with our text embeddings to create a shared semantic space. The encoder weights of AION-1-Base and \texttt{text-embedding-3-large} are not updated during training.

Our alignment architecture uses four-layer residual multilayer perceptrons (MLPs) as projection heads for both image and text, learning a shared 1024-dimensional embedding space through a contrastive loss. Motivated by prior work suggesting that shorter captions may lead to higher performance in contrastive learning~\citep{li2023an}, we compare training on the original multi-paragraph descriptions against single-sentence summaries generated by GPT-4.1-nano (see Appendix Fig.~\ref{fig:summary_prompt} for prompt). We train on a single NVIDIA A100 GPU using AdamW optimizer~\citep{loshchilov2018decoupled} with learning rate $10^{-4}$, weight decay 0.05, and cosine annealing. We leverage AION-1-Base encoder embeddings averaged over token-space (768-dimensional) for images and \texttt{text-embedding-3-large} (3072-dimensional) for text. This average pooling follows the AION embedding protocol for similarity search and morphology classification. We use symmetric cross-entropy~\citep[InfoNCE]{infonce} as the contrastive loss.
\subsection{Evaluating search capabilities}\label{sec:evaluating_search}
We adopt the retrieval evaluation protocol established for AION, assessing performance on three categories with varying rarity in the AION benchmark datasets of Legacy Survey images: spiral galaxies (43,793 in GZ-DECaLS dataset, making up 26\% of the dataset being searched), mergers (4,036 in GZ-DECaLS dataset, 2\%), and gravitational lenses (770 in HSC strong lens catalog cross-matched with Legacy Survey images, 0.1\%). Additional details about dataset construction are described in the Appendix~\ref{sec:retrieval_data}. Each image has relevance scores ranging from 0.0 to 1.0: for spirals and mergers, the score represents the fraction of human volunteers who identified each category; for lenses, the images receive 1.0 if present in expert-curated lens catalogs and 0.0 otherwise. The rarity of gravitational lenses, with only a few thousand confident examples in the literature~\citep{lenscat}, coupled with their usefulness in constraining cosmology and measuring the mass and shape of dark matter halos, makes them a challenging and scientifically valuable target.

To measure the search performance, we use the Discounted Cumulative Gain (DCG) to evaluate ranking quality, capturing both precision and the relative ordering of retrieved results. We normalize DCG by the ideal (sorted) ordering to obtain nDCG. If $r_i$ is the relevance label of the candidate ranked at position $i$, the discounted cumulative gain of the top-10 images is,
\begin{equation}
\mathrm{DCG} @ 10=\sum_{i=1}^{10} \frac{2^{r_i}-1}{\log _2(i+1)}\,.
\end{equation}
We evaluate against unsupervised baseline models from astronomy (AION-1 variants, AstroCLIP; \citealt{AstroCLip}, and \citealt{stein}) and computer vision (DINOv2; \citealt{dinov2}). These baselines perform retrieval by computing cosine similarity between query and candidate image embeddings. Queries are high-confidence examples ($>$90\,\% volunteer agreement for morphologies, or presence in lens catalogs), and performance is measured by averaging nDCG@10 across queries. For our model, we query with the text embeddings for: ``visible spiral arms'', ``merging'', and ``gravitational lens''. We note that retrieval performance depends on query phrasing, for example ``spiral galaxy with visible arms'' and ``galaxy showing spiral structure'' can yield different rankings, and a systematic exploration of query sensitivity is left to future work.
\subsection{Post-search VLM re-ranking}\label{sec:reranking}
Re-ranking is a retrieval strategy in which a fast embedding search retrieves a small candidate set, and a more expensive model then refines its ordering~\citep{nogueira2019passage}. This can improve performance because the first stage is optimized for broad scalable retrieval, while the re-ranker performs a more expensive evaluation of each candidate. For image search, a VLM re-ranker examines each candidate using the user's query as the prompt, effectively acting as a query-specific classifier over the retrieved set (e.g., \citealt{inquire}).

To improve the accuracy of a search, we experiment with a re-ranking step where a VLM examines the top-k retrieved images and assigns a score (1-10) based on the search query (e.g., ``Does this galaxy image display signs of gravitational lensing? Rank 1-10.''), and reorders based on these scores. Specifically, for each retrieval task (spirals, mergers, and lenses) conducted in \S\ref{sec:evaluating_search}, we apply GPT-4.1 to score each of the top-1000 retrieved images and reorder them based on these scores. This automated verification step is inspired by current discovery pipelines, where classifiers produce thousands of false positives that require manual inspection by expert teams (e.g.~\citealt{holismokes} for lenses).

To push this approach further, we investigate whether scaling compute can improve retrieval results. To evaluate this hypothesis, we construct a controlled test dataset using higher-resolution HSC images, the same data used for initial human identification, rather than lower-resolution Legacy Survey images, comprising 20,000 non-lenses and 200 confirmed gravitational lenses from published HSC strong-lens catalogs (see Appendix~\ref{sec:retrieval_data}).

\begin{figure*}[t]
    \centering
    \includegraphics[width=0.9\linewidth]{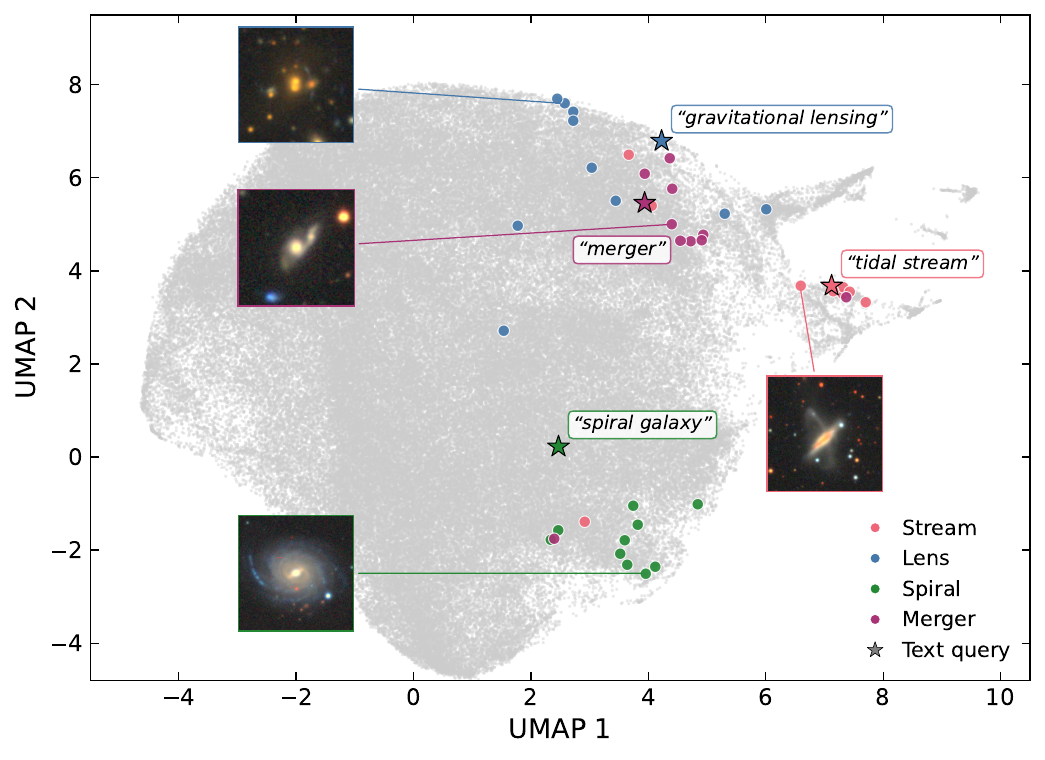}
    \caption{UMAP projection of the AION-Search embedding space shows text queries falling near their corresponding image clusters. Grey points are the training set images and colored points are held-out examples of spirals, mergers, lenses, and streams drawn from the retrieval benchmarks (\S\ref{sec:evaluating_search}) and stream candidates (\S\ref{sec:streams}). Stars mark text query embeddings. UMAP preserves local neighborhood structure but not global distances, so the global layout is for illustrative purposes only.}
    \label{fig:umap}
\end{figure*}
\begin{figure*}[t]
    \centering
    \includegraphics[width=\linewidth]{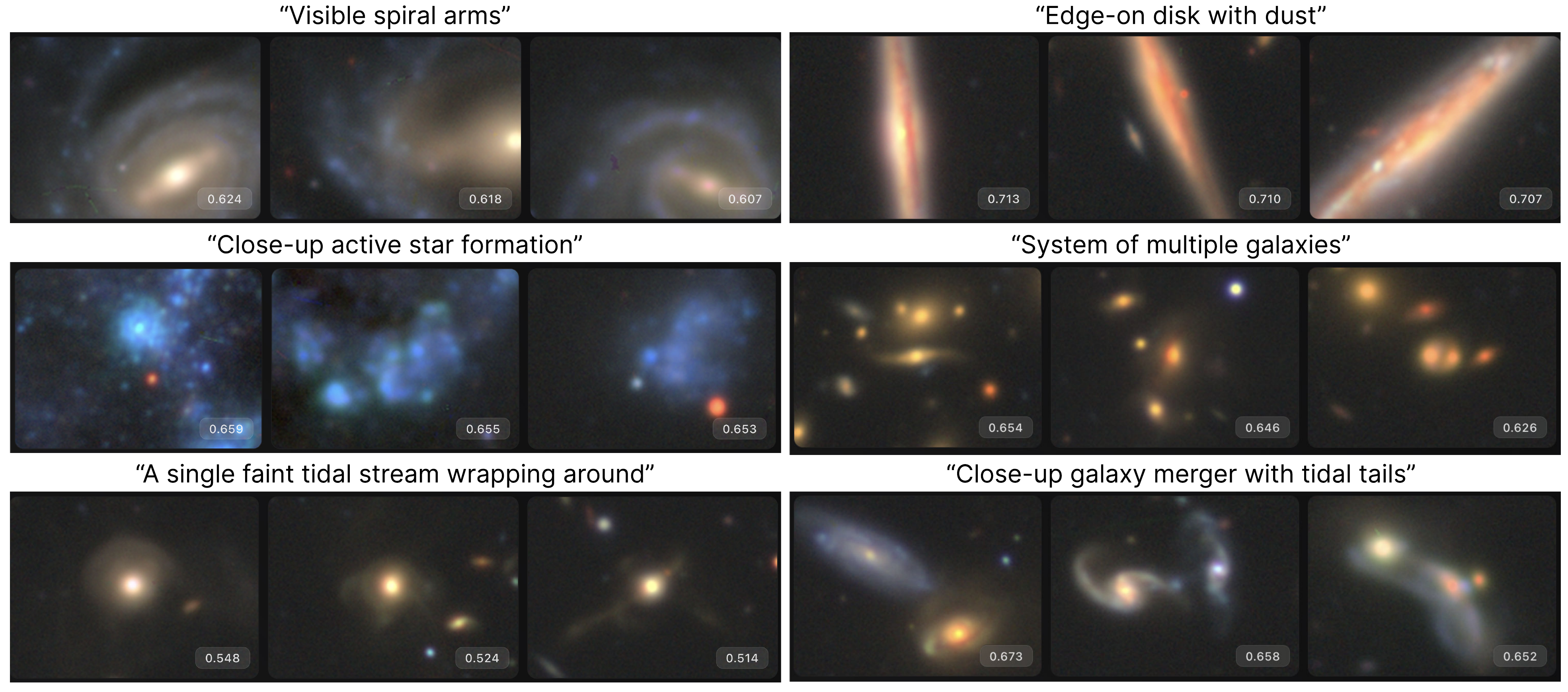}
  \caption{AION-Search enables semantic retrieval of galaxies matching free-form natural language queries. Top three retrieved images from the HSC survey for each query demonstrate the system's ability to identify specific astronomical phenomena that would traditionally require volunteer labeling to catalog and train a supervised classifier. Cosine similarity scores are shown in the bottom right of each image.}
\label{fig:example_searches}
\end{figure*}
\begin{figure*}[t]
    \centering
    \includegraphics[width=\linewidth]{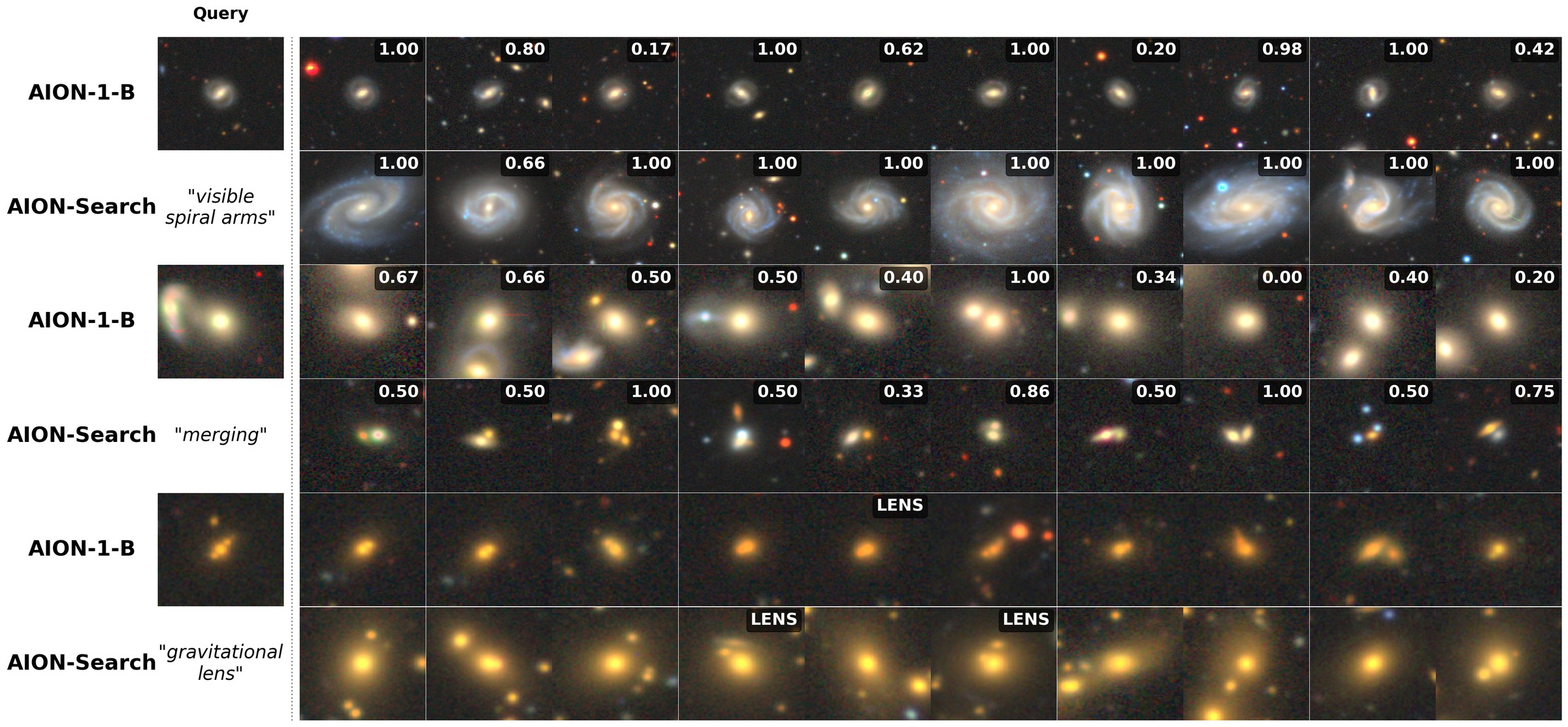}
    \caption{AION-Search text-query retrieval returns more semantically relevant matches than AION-1-B image-query similarity search. Top-10 retrieved images for spirals, mergers, and gravitational lenses for each method. For AION-1-B, each row shows the result from a single representative image query, one of many used to compute the averaged retrieval scores in Table~\ref{tab:spiral_merger_lens}. Galaxy Zoo vote fractions are shown above each result. ``LENS'' labels mark confirmed gravitational lenses from published catalogs.}
    \label{fig:demo_comparison}
\end{figure*}
We perform semantic search using AION-Search with the query ``gravitational lens'' and measure performance by counting confirmed lenses in the top-100 results after re-ranking the initial top-1000 retrievals, comparing against two baselines: the original AION-Search ranking and AION-1-B similarity search. Our experimental design investigates two dimensions of computational scaling: (1) model size by comparing GPT-4.1-nano, GPT-4.1-mini, and GPT-4.1 in ascending order of size and cost; and (2) inference-time strategy through $n$-sample averaging, where we generate $n$ independent scores per image and average them for the final ranking.
\subsection{Zero-shot classification on Galaxy10}\label{sec:gz10}
For galaxy morphology classification, supervised classifiers are currently the standard approach when labeled examples are available~\citep{dieleman2015,2022walmsley,walmsley2023gzdesi}. We therefore compare AION-Search against supervised classifiers using Galaxy10 DECaLS~\citep{leung2024galaxy10}, which provides a useful benchmark for testing whether AION-Search can perform zero-shot morphology classification. Galaxy10 is a dataset of galaxy images from the DESI Legacy Imaging Surveys~\citep{legacy} classified into ten morphology classes using Galaxy Zoo volunteer votes~\citep{lintott_galaxyzoo,2022walmsley} and has been previously used to evaluate VLM-generated galaxy descriptions~\citep{zaman2025astrollavaunificationastronomicaldata}. We adopt the same 796-galaxy test set used in~\citet{2025aion1} Table~2 for direct comparison with supervised baselines.

For AION-Search, we perform zero-shot classification by embedding all 796 images and assigning each image to the class with the highest cosine similarity between image and class text label. To test whether contrastive training improves or degrades the underlying AION representations, we also train a two-layer MLP head on frozen AION-Search embeddings using the same protocol as~\citet{2025aion1}. We also evaluate VLM description accuracy on this benchmark by generating descriptions for all 796 images with GPT-4.1-mini, GPT-5-mini, and Qwen3-VL-235B~\citep{bai2025qwen3vltechnicalreport}, with the prompt in Fig.~\ref{fig:prompt}, then classifying each description into one of the ten classes using Gemini-2.5-Flash as a judge.
\subsection{Deployment}\label{sec:deployment}
To make AION-Search available for interactive exploration, we develop a publicly accessible web application. We generate AION-Search embeddings for nearly all Legacy Survey DR10 galaxies in MMU, over 100 million in total, which we release on Hugging Face. For the public search application, we index the 19 million galaxies brighter than $r = 21$\,mag using Zilliz\footnote{\url{https://zilliz.com}}, a managed open-source vector search service. This threshold balances retrieval coverage against vector database storage cost, and retrieval quality for galaxies near the faint limit has not been separately validated, as the training set uses $r < 19$\,mag. The search interface is hosted as a web application on Hugging Face Spaces\footnote{\url{https://astronolan-aion-search.hf.space/}}.

The application supports three query modes. In text search, the user enters a natural-language query (e.g., ``galaxy with tidal stream'') that is embedded and compared against all catalog embeddings via cosine similarity. In image similarity search, the user provides a reference galaxy image and the system retrieves the most similar objects in embedding space. Hybrid search combines a reference image with a text query via vector addition in the embedding space, with adjustable weighting between the two.

The application has been used for both outreach and research, accumulating over 1,000 queries to date.
\section{Results}\label{sec:results}
We find that VLM descriptions provide a sufficient training signal for contrastive learning to enable text-based retrieval of galaxy images.
\subsection{VLM image descriptions}\label{sec:gz_results}
We first analyze whether current VLMs from OpenAI~\citep{openai_gpt41,openai_o3_systemcard}, Google~\citep{comanici2025gemini25pushingfrontier}, and Alibaba\footnote{Alibaba Qwen models were accessed through OpenRouter: \url{https://openrouter.ai}}~\citep{bai2025qwen3vltechnicalreport} can accurately describe the visual properties of galaxy images using our curated Galaxy Zoo test set (\S\ref{sec:gz_curation}).

We evaluate VLM performance using the two settings described in \S\ref{sec:gz_curation}. First, we test a single strong model, OpenAI o3, to see how well it answers Galaxy Zoo questions when the questions are explicitly provided. Second, we test the ability of all models to generate free-form descriptions without access to the Galaxy Zoo questions, aligning this setup with the procedure used for creating our training data.

In the direct-question setting, we quantify the fraction of images where model classifications exactly match the majority of volunteers. An image is counted as correct only if the model reproduces the full majority-vote Galaxy Zoo decision-tree path. Under this strict criteria, o3 achieves $23.6\pm1.0$\% of galaxy classifications exactly matching over three runs. While this is lower than the average volunteer ($51.9\pm 2.4$\%), it significantly exceeds random chance ($4.1\pm1.1$\%), demonstrating that VLMs can provide a noisy but informative signal for human labeling tasks. We compute the average volunteer rate by, for each galaxy, multiplying the majority vote fractions along the decision path and then averaging across galaxies, treating the majority vote as ground truth.

In the second evaluation, we assess free-form descriptions generated by the VLMs by measuring the fraction of Galaxy Zoo decision-tree answers that match the volunteer-derived labels. For every image, each model first generates a free-form description, which is subsequently converted by a judge model into Galaxy Zoo decision-tree answers. The percentage of individual answers that match the Galaxy Zoo label is shown in Figure~\ref{fig:vlm_description_bench}. Evaluating 14 different models from Google and OpenAI with three runs each, GPT-4.1-mini was optimal for our budget constraints, with $50.1\pm3.3$\% of individual Galaxy Zoo answers matching the volunteer-derived labels, while models like o3 offer stronger performance but higher cost.

We developed our prompt through a combination of manual tweaking and automated prompt optimization~\citep{aide2024}. We designed the prompt to remain general enough to capture galaxy features beyond Galaxy Zoo's fixed question set (see Fig.~\ref{fig:prompt} for prompt). We use this setup for CLIP training set caption generation.

\subsection{Search performance}
Figure~\ref{fig:umap} shows a two-dimensional UMAP~\citep{mcinnes2020umap} projection of the AION-Search embedding space, fitted on the training set. We project held-out image examples of spirals, mergers, lenses, and streams drawn from the retrieval benchmarks (\S\ref{sec:evaluating_search}) and stream candidates (\S\ref{sec:streams}), along with text query embeddings for ``spiral galaxy'', ``merger'', ``gravitational lensing'', and ``tidal stream''. The text embeddings fall near their corresponding image clusters, confirming that the contrastive training produces a shared space where text and image modalities are aligned.

We demonstrate AION-Search by retrieving galaxies for complex queries like ``close-up active star formation'' and ``a single faint tidal stream'' in Figure~\ref{fig:example_searches}, which are free-form searches that traditional supervised methods cannot perform and would typically require months-long volunteer labeling campaigns to enable. Using our retrieval benchmarks, Figure~\ref{fig:demo_comparison} illustrates the qualitative difference between AION-1-B similarity search and AION-Search text-query retrieval across all three categories. AION-Search returns relevant results, while AION-1-B returns visually similar but sometimes irrelevant images.

For a quantitative analysis of retrieval performance, Table~\ref{tab:spiral_merger_lens} reports nDCG@10 scores for retrieval for spiral galaxies, mergers, and gravitational lenses. Our AION-Search model, trained on the summarized VLM-generated captions, consistently outperforms similarity-based baselines across all categories. For spirals, we observe an nDCG@10 of 0.941 compared to 0.643 for the strongest baseline (AION-1-L). For mergers, the model achieves 0.554 versus 0.384 (AION-1-XL), and for gravitational lenses 0.173 compared to 0.015 (AION-1-XL).

Baseline methods using visual similarity to find lenses return fewer than a single lens on average in their top-10 results when averaged across multiple queries. In contrast, our semantic approach retrieves two lenses in the top-10. To validate, we conducted a 10-fold cross-validation analysis (Appendix~\ref{sec:retrieval_data}, Table~\ref{tab:kfold_results}), which confirms that AION-Search consistently outperforms AION-1-B across all categories, though lens retrieval exhibits high variance due to the rarity of positives and our use of a single text query. Some retrieved candidates not counted as lenses under the initial protocol appear in other lens catalogs that were not used, so we conduct a sensitivity analysis varying the lens catalog set and cross-match radius (Appendix~\ref{sec:lens_sensitivity}, Table~\ref{tab:lens_sensitivity}), which shows that these evaluation choices do not affect the relative improvement of AION-Search compared to image similarity search.

Training on single-sentence summaries improves nDCG@10 by 0.143 (spirals), 0.084 (mergers), and 0.173 (lenses) relative to multi-paragraph descriptions on the full evaluation set, consistent with prior work suggesting that shorter captions reduce noise and overfitting to caption-specific details~\citep{li2023an}. However, 10-fold cross-validation shows a more mixed picture: 0.066 (spirals), -0.061 (mergers), and 0.051 (lenses) for summaries. We therefore do not find clear evidence that training on summaries reliably improves retrieval across categories.

\begin{table}[t]
\centering
\footnotesize
\begin{tabular}{lccc}
\toprule
& Spirals & Mergers & Lenses \\
\midrule
AION-1-B & 0.632 & 0.281 & 0.012 \\
AION-1-L & 0.643 & 0.303 & 0.011 \\
AION-1-XL & 0.621 & 0.384 & 0.015 \\
AstroCLIP~\citep{AstroCLip} & 0.602 & 0.248 & 0.006 \\
\citet{stein} & 0.590 & 0.340 & 0.007 \\
DinoV2~\citep{dinov2} & 0.477 & 0.060 & 0.003 \\
Random & 0.263 & 0.037 & 0.000 \\
\midrule
AION-Search & 0.941  & 0.554 & 0.173 \\
AION-Search (re-rank) & 0.992  & 0.678 & 0.328 \\
\bottomrule
\end{tabular}
\caption{Semantic search performance via nDCG@10 scores on astronomical images demonstrates superiority over similarity-based methods. AION-Search, trained on VLM-generated captions, achieves strong performance using text queries (``visible spiral arms'', ``merging'', ``gravitational lens'') compared to similarity-based baselines using image queries. Performance gains are most pronounced for rare phenomena. GPT-4.1 re-ranking of top-1000 results further improves performance across all categories. Baseline methods include various AION-1 model sizes (B, L, XL), AstroCLIP~\citep{AstroCLip}, astronomical self-supervised models~\citep{stein}, and DINOv2~\citep{dinov2}. Retrieval scores for randomly shuffled data are shown for reference. See Table~\ref{tab:kfold_results} for 10-fold cross-validation uncertainty estimates on AION-1-B and AION-Search.}
\label{tab:spiral_merger_lens}
\end{table}

\subsection{Re-ranking}
On our HSC dataset of 200 confirmed gravitational lenses hidden among 20,000 non-lenses, AION-Search successfully retrieves 38 lenses within its top-1000 retrievals, demonstrating strong initial recall. While the top-100 precision matched AION-1-B one-shot similarity search (7 lenses), we find that VLM re-ranking can greatly improve performance. By applying GPT-4.1 with 5-sample averaging to re-rank the initial 1000 retrievals, we nearly doubled the number of gravitational lenses in the top-100, increasing from 7 to 13.

\begin{figure}[t]
  \centering
  \begin{minipage}[t]{0.48\textwidth}
    \vspace{27pt}
    \centering
    \includegraphics[width=\linewidth]{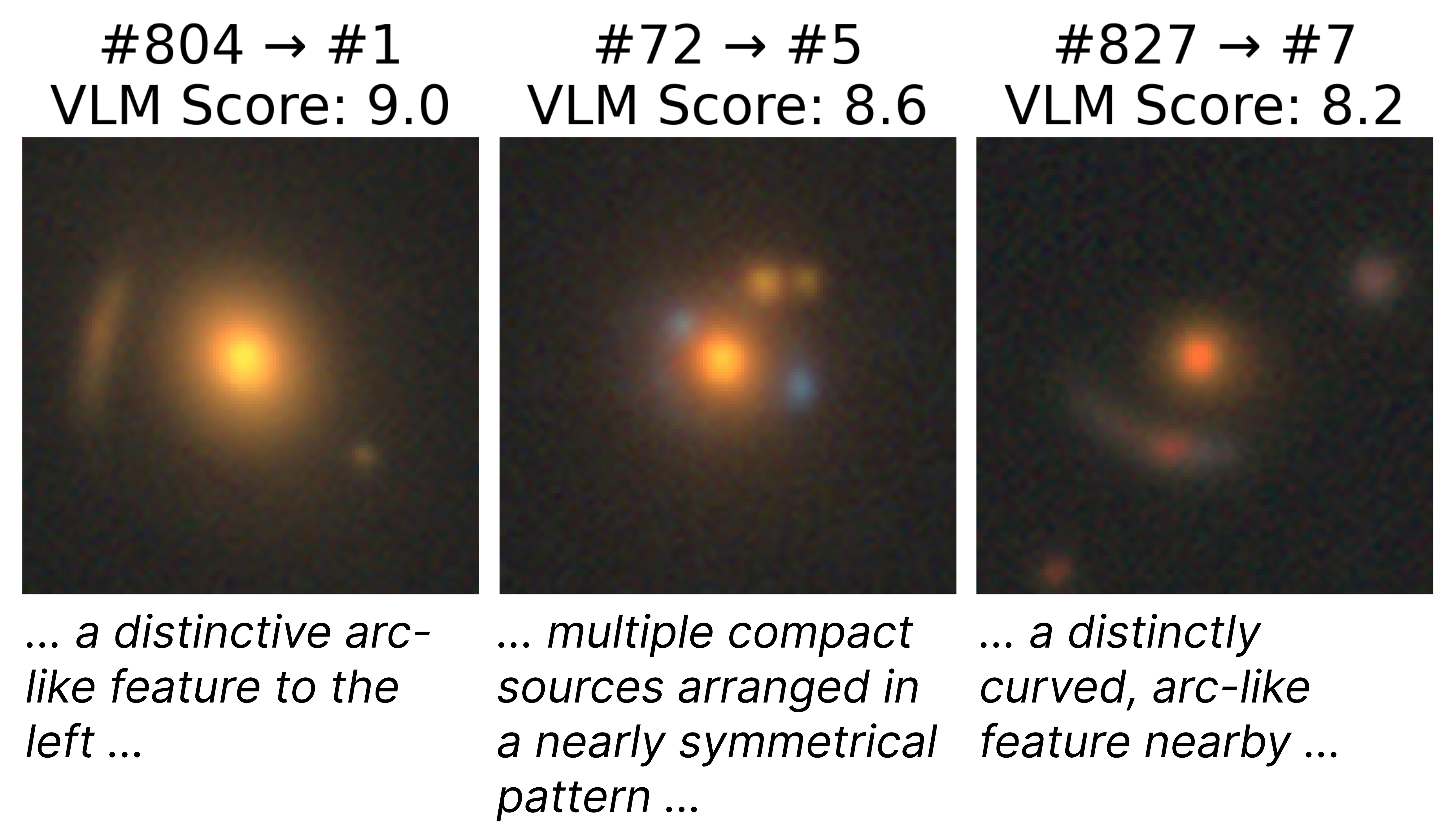}
    \vspace{0.5em}
  \end{minipage}
  \hfill
  \begin{minipage}[t]{0.48\textwidth}
    \vspace{0pt}
    \centering
    \includegraphics[width=\linewidth]{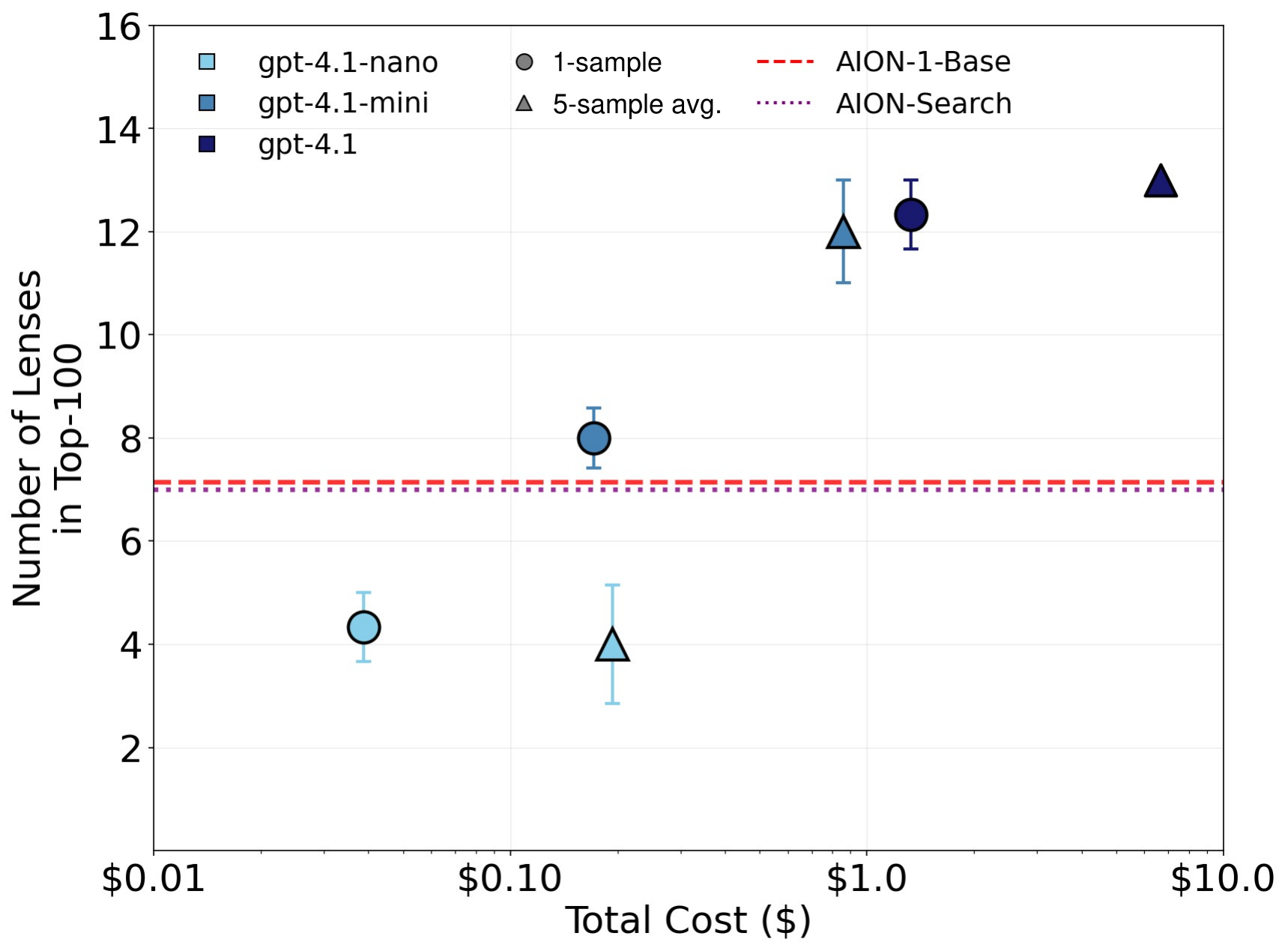}
    \vspace{0.5em}
  \end{minipage}
  \caption{VLM re-ranking improves gravitational lens discovery and performance scales with compute.
\textit{Top:} First three gravitational lenses identified by GPT-4.1 with 5-sample averaging for re-ranking, showing rank improvements from initial AION-Search results along with excerpts from the GPT-4.1 explanations (see Appendix Fig.~\ref{fig:rerank_prompt} for prompt).
\textit{Bottom:} Performance scales with compute: number of confirmed lenses in top-100 results after re-ranking the top-1000 AION-Search results. Baseline methods (AION similarity search and no re-ranking) find only seven lenses. VLM re-ranking performance improves with both model size (GPT-4.1-nano to GPT-4.1) and inference-time compute via N-sample averaging, reaching 13 lenses with GPT-4.1 5-sample averaging. This demonstrates that spending more compute at search-time improves discovery of rare astronomical phenomena.}
  \label{fig:reranking}
\end{figure}

Furthermore, our results reveal a relationship between compute in re-ranking and retrieval performance for rare astronomical phenomena (Figure~\ref{fig:reranking}). As we scale from GPT-4.1-nano through GPT-4.1, and from single evaluations to 5-sample averaging, we increase the number of retrieved lenses. Applying GPT-4.1 re-ranking to the full retrieval evaluation set improves nDCG@10 across all three categories (Table~\ref{tab:spiral_merger_lens}).
\begin{figure*}[t]
    \centering
    \includegraphics[width=\linewidth]{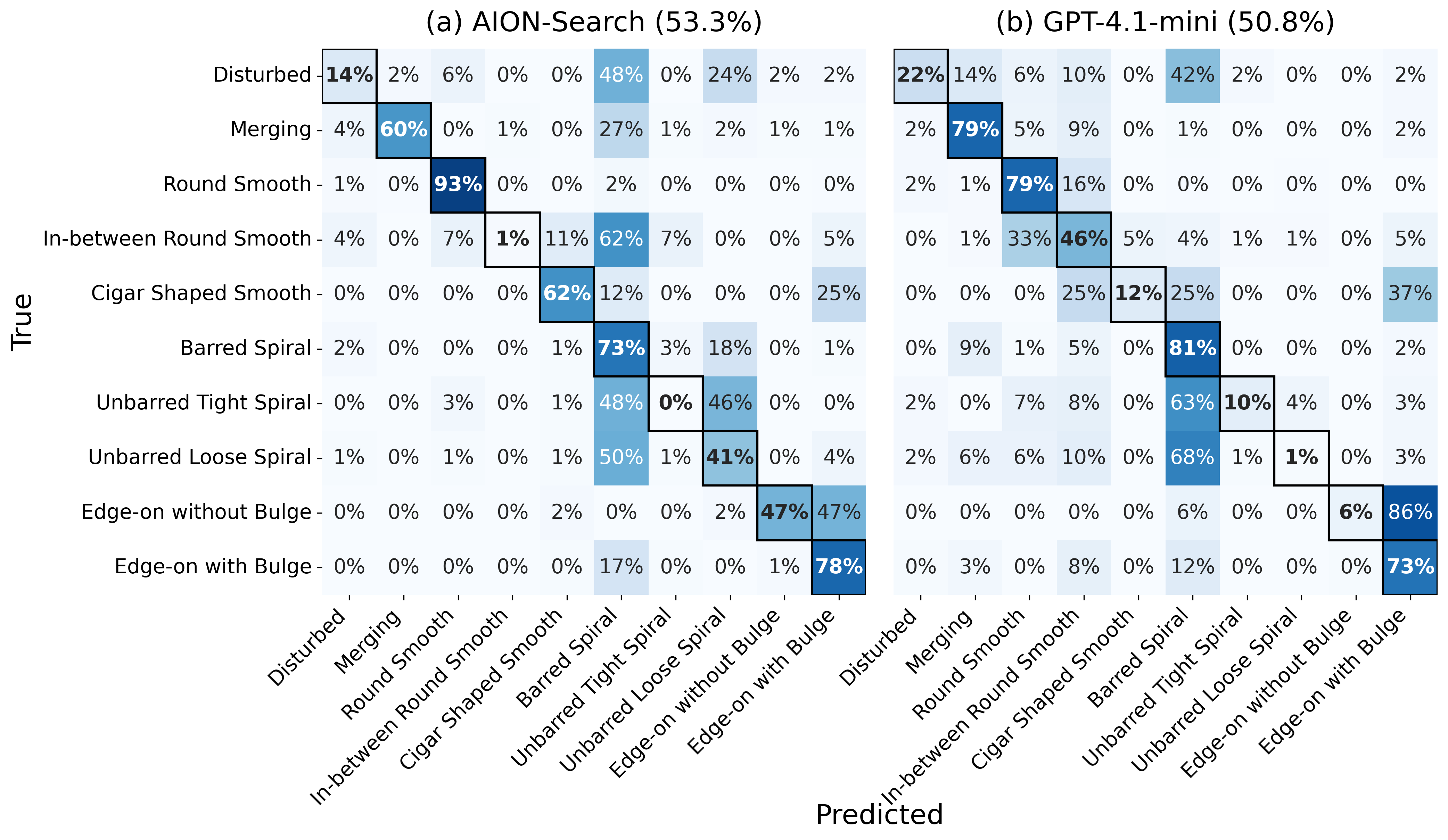}
    \caption{Zero-shot Galaxy10 classification reveals strong broad-class separation but poor fine-grained performance, with a barred spiral over-prediction in both methods. (a) AION-Search assigns each image to the class with the highest cosine similarity between its embedding and the class label embedding. (b) GPT-4.1-mini generates a free-form description that is then classified by a Gemini-2.5-Flash judge. Each cell shows the percentage of true class instances classified as the predicted class.}
    \label{fig:gz10_confusion}
\end{figure*}
\subsection{Zero-shot classification}\label{sec:gz10_results}
On Galaxy10 (\S\ref{sec:gz10}), AION-Search zero-shot classification achieves 53.3\%, slightly exceeding the accuracy of the model that generated its training data (GPT-4.1-mini at 50.8\%). Table~\ref{tab:gz10} shows that AION-Search with MLP probing achieves 84.1\%, matching AION-1-B (84.0\%) in the same conditions. This could indicate that the morphological information is preserved in the vision-only representations, and that the language alignment transforms the space to be text-accessible without destroying or enriching the required information for this task.

For the other VLMs, GPT-5-mini achieves 56.0\% and Qwen3-VL-235B achieves 47.1\%. Being an open model, Qwen3-VL has the advantage of local execution, but its performance remains behind closed models for this task.

Inspection of the confusion matrices (Figure~\ref{fig:gz10_confusion}) reveals that GPT-4.1-mini strongly over-predicts barred spirals, the most significant source of error. To investigate, we ran the same pipeline blind, without providing the actual images to the API. GPT-4.1-mini did not refuse and still produced descriptions classifying 75.4\% as barred spirals and achieving 11.2\% accuracy, close to random as expected.

This indicates that the over-prediction is not driven solely by visual evidence in the images, but also reflects a prior in the model or prompt toward barred-spiral descriptions. We do not attempt to determine the origin of this prior here. This over-prediction remains even after excluding the word ``bars'' from our prompt (see prompt in Fig.~\ref{fig:prompt}). AION-Search inherits the barred-spiral over-prediction from its training descriptions (Figure~\ref{fig:gz10_confusion}a). Overall, broad classes (smooth, spiral, edge-on) are well classified, while fine-grained distinctions (in-between smooth vs.\ cigar-shaped smooth) produce more confusion.

For reference, author N.K.\ manually classified the same 796 test images, achieving 90.1\% accuracy, first assigning each galaxy to a broad category, then performing fine-grained sub-category classification. This suggests that $\sim$90\% is likely the best achievable performance on this classification task, given the inherent ambiguity in fine-grained galaxy morphology labels.

\begin{table}[t]
\centering
\footnotesize
\begin{tabular}{lc}
\toprule
Model & Accuracy (\%) \\
\midrule
EfficientNet~\citep{tan2019efficientnet} & 80.0 \\
DINOv2 + MLP probe~\citep{dinov2} & 71.4 \\
AION-1-B + MLP probe~\citep{2025aion1} & 84.0 \\
AION-1-XL + MLP probe~\citep{2025aion1} & 86.5 \\
AION-1-L + MLP probe~\citep{2025aion1} & 87.2 \\
ZooBot + MLP probe~\citep{2022walmsley} & 89.6 \\
Human & 90.1 \\
\midrule
AION-Search (zero-shot) & 53.3 \\
AION-Search + MLP probe & 84.1 \\
GPT-4.1-mini (zero-shot + judge) & 50.8 \\
\bottomrule
\end{tabular}
\caption{Galaxy10 classification accuracy for supervised and zero-shot methods. Supervised baselines (top) are from \citet{2025aion1}, using MLP (multilayer perceptron) heads on frozen embeddings except EfficientNet (trained end-to-end). Our methods (bottom) include both supervised and zero-shot evaluation. AION-Search + MLP probe matches AION-1-B, confirming that contrastive alignment does not degrade morphological information. As expected, AION-Search zero-shot classification is close to the performance of the GPT-4.1-mini that generated its training descriptions.}
\label{tab:gz10}
\end{table}

\subsection{Discovery of new extragalactic streams}\label{sec:streams}
Beyond benchmarks, we test whether AION-Search can support exploratory discovery by using it to search for extragalactic stellar streams. Stellar streams form when globular clusters or dwarf satellite galaxies are tidally disrupted by their host galaxy, producing elongated, low-surface-brightness tidal debris that traces the gravitational potential of the host~\citep{Johnston1999,Helmi1999,Delgado2010,shipp2021,koposov2023,Nibauer_2023}. Dwarf galaxy streams are more massive and more extended than those produced by globular clusters, making them detectable in ground-based imaging such as the Legacy Survey. Notably, extragalactic stellar streams can provide a powerful probe of the shape of dark matter halos \citep{Nibauer_2023}. Crucially, AION-Search was not explicitly trained to find streams, as its training set consists of randomly selected galaxy images with general-purpose VLM descriptions. The results presented here therefore reflect generalization for a scientific application without using labeling campaigns.

\subsubsection{Stream discovery via search}
\begin{figure}[t]
    \centering
    \includegraphics[width=\linewidth]{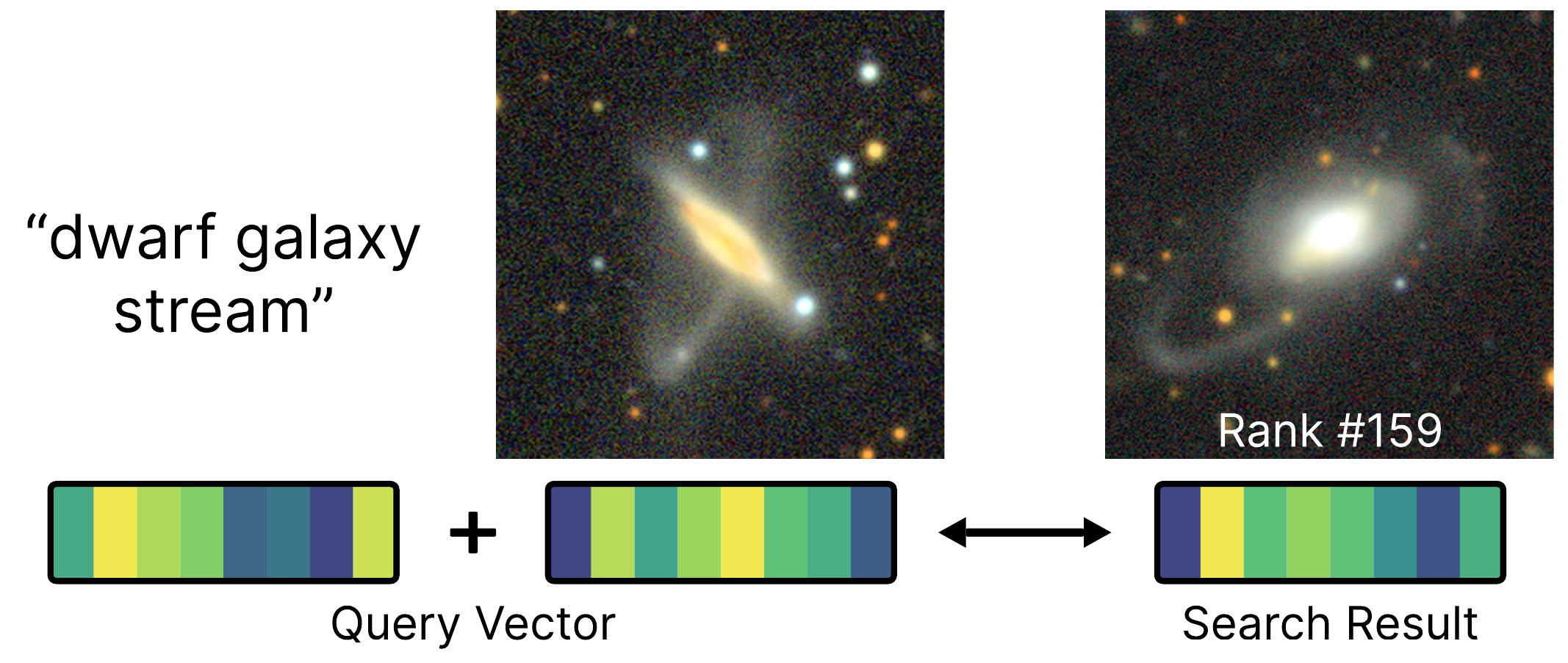}
    \caption{Hybrid image and text search enhances the capability for discovery of galaxies with specific features. The text query ``dwarf galaxy stream'' is combined with the image embedding of a galaxy at $(\alpha, \delta) = (203.196^\circ, -36.541^\circ)$ via weighted vector addition in the shared embedding space. The result at rank 159 displays a clear tidal stream, demonstrating the utility of combining visual similarity with semantic descriptions.}
    \label{fig:hybrid_search_example}
\end{figure}

Using the publicly available web application described in \S\ref{sec:deployment}, we searched for extragalactic streams using a strategy combining text and image queries. Beginning with the text query ``galaxy with stream,'' we identified two promising stream candidates within the top-100 results. We then used these discoveries as seeds for image + text searches: the embedding of each newly found stream image was added to the text embedding for queries such as ``galaxy with stream,'' ``tidal streams,'' and ``tidal shells,'' with adjustable weighting between the image and text contributions. Figure~\ref{fig:hybrid_search_example} illustrates an example of image + text search: combining the text query ``dwarf galaxy stream'' with the image embedding of a galaxy returns a galaxy at rank 159 which displays a tidal stream. Each new stream candidate was in turn used as a seed for further searches, progressively building a catalog of candidates through this exploration process.

\begin{figure*}[t]
    \centering
    \includegraphics[width=\linewidth]{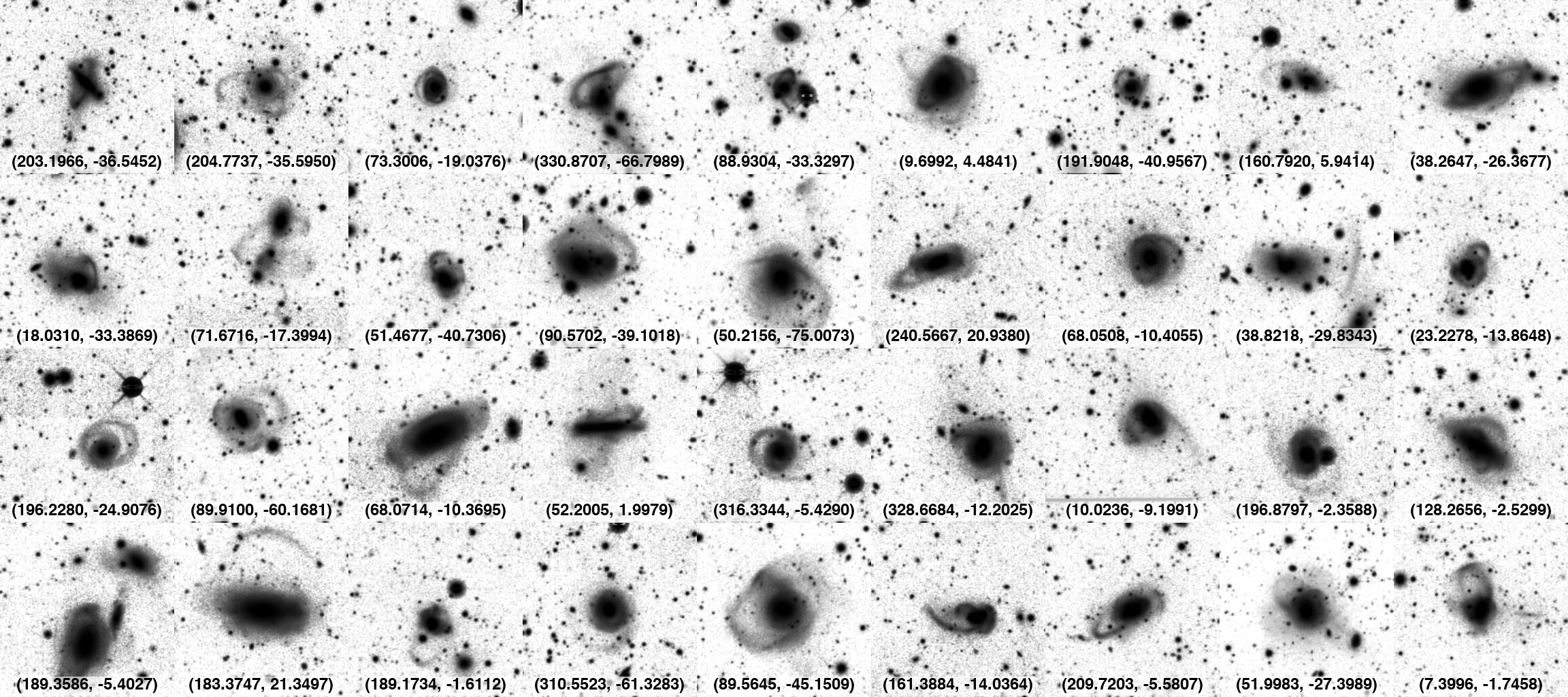}
    \caption{36 new extragalactic stellar stream candidates discovered using AION-Search. Each panel shows a Legacy Survey DR10 cutout with an arcsinh stretch to enhance low surface brightness features, labeled with RA and Dec. Candidates were found through iterative hybrid image and text queries (\S\ref{sec:streams}) and confirmed as novel by cross-matching against 11 published catalogs of tidal features.}
    \label{fig:stream_collage}
\end{figure*}

After inspecting thousands of search results from dozens of searches, this procedure yielded 36 new extragalactic stream candidates (Figure~\ref{fig:stream_collage}).
\subsubsection{Cross-match with existing catalogs}
To assess the novelty of these candidates, we compiled a reference set of 12,718 galaxies with identified streams or tidal features from the following published catalogs: \citet{atkinson2013}, \citet{hood2018}, \citet{paudel2018}, \citet{martinezdelgado2023}, \citet{mirocarretero2023,mirocarretero2024}, \citet{skryabina2024}, \citet{pippert2025}, \citet{sola2025a,sola2025b}, and~\citet{desmons2025}. We note that the majority of objects in these catalogs are tidal features broadly defined (including shells, tails, and asymmetric halos) and that catalogs of confirmed extragalactic streams typically only contain a few dozen objects each. Notably, \citet{sola2025a} and~\citet{sola2025b} use the same Legacy Survey DR10 imaging data as our deployed AION-Search index.

We cross-matched the positions of our 36 stream candidates against all objects in the combined reference catalog using a $30''$ match radius. None of our candidates matched any previously cataloged stream or tidal feature. This may reflect limited overlap with existing catalogs, many of which restrict their samples to nearby galaxies whose tidal features extend beyond the AION-1 crop. For candidates that do fall in the space of existing catalogs, this suggests AION-Search can identify streams missed by methods relying on supervised classifiers or visual inspection campaigns.

\section{Discussion and Conclusions}\label{sec:discussion}
Our results demonstrate that VLMs can both generate descriptions for semantic search and score images to enable effective re-ranking of search results. Although VLM descriptions contain hallucinations (e.g., saying a galaxy has two spiral arms instead of three like in Fig.~\ref{fig:vlm_description_bench}), these imperfect annotations nonetheless provide an effective training signal for contrastive learning. Our resulting semantic search model achieves strong zero-shot performance with text queries for spiral galaxies, mergers, and gravitational lenses, substantially outperforming similarity-based retrieval methods that rely on example images. 

VLM-based re-ranking consistently improves retrieval performance across all tasks (Table~\ref{tab:spiral_merger_lens}): spirals improve from 0.941 to 0.992 nDCG@10 and mergers from 0.554 to 0.678, improvements even when AION-Search already achieves strong baseline performance. The most substantial gains occur for gravitational lenses, where re-ranking nearly doubles the number found in the top-100 (from 7 to 13), with performance scaling with compute. This pattern makes re-ranking particularly valuable for discovering rare astronomical phenomena at scale, offering an alternative to resource-intensive manual labeling campaigns that typically require months to years of volunteer effort~\citep{walmsley2023gzdesi,euclidcollaboration2025euclidquickdatarelease,lintott_galaxyzoo}.

Re-ranking could be extended through agentic approaches that provide models with image manipulation capabilities and domain-specific computational tools, such as astrophysical modeling packages (e.g., lens modeling with the \texttt{lenstronomy} package; \citealt{lenstronomy}). Recent work demonstrates that tool-augmented multimodal reasoning substantially improves visual understanding tasks~\citep{shao2024visual, hu2024visual, yang2023mmreactpromptingchatgptmultimodal}. If a discovery is made, the analysis pipeline produced by the agent can be examined and verified by a human expert. Such methods could provide an effective way to scale compute beyond the N-sample averaging explored in this work.

The minimal training required for AION-Search demonstrates the quality of the AION representations. Space-based surveys such as Euclid and Roman will provide high-resolution imaging for hundreds of millions to billions of galaxies with detailed resolved morphology, making semantic search increasingly valuable~\citep{euclidredbook,roman}. The AION architecture addresses this need through its unified latent space that handles multiple modalities (four-band Legacy Survey and five-band HSC images as demonstrated here, plus spectroscopic measurements of stars and galaxies), allowing the model to be adapted to new telescope data. Future implementations of AION-Search could incorporate time-series photometry or spectroscopic observations of stars and other astronomical sources, providing a single general-purpose semantic search engine for astronomy.

Several limitations constrain our approach. VLMs often fail to describe fine-grained features, and our model inherits biases from GPT-4.1-mini. VLM descriptions are also sometimes ambiguous between classes (e.g., describing a galaxy as ``disturbed or merging''), making classification noisy and limiting the precision of any benchmark that assigns a single label. Furthermore, the AION-1 encoder generates embeddings from $96\times96$ pixel crops of the $160\times160$ pixel cutouts~\citep{2025aion1}, which can truncate extended features for nearby galaxies.

Both of our VLM benchmarks have limitations. Our benchmark in \S\ref{sec:gz_curation} uses a decision-tree question structure where absent information is scored as incorrect: if a spiral galaxy does not have a bar, a good description may simply not mention bars, but the benchmark penalizes this omission. The decision tree removes most such cases (questions about spiral features are not asked for smooth galaxies), but some remain. Galaxy10 provides a larger test set but with coarser labels. A more informative benchmark would consist of a few hundred galaxies with expert-written rubrics describing all notable features, covering not just common morphologies but also rarer phenomena such as AGN, tidal streams, and gravitational lenses. We leave the construction of such a benchmark to future work.

The source of the astronomy knowledge in general-purpose VLMs is difficult to determine, especially for closed models. Public astronomy resources such as Galaxy Zoo~\citep{lintott_galaxyzoo} could appear in training corpora, but we cannot quantify the contribution of any particular source. We therefore interpret our results as an empirical evaluation of the tested VLMs, rather than evidence that Galaxy Zoo or any other specific pretraining source is responsible for their performance.

Caption quality may depend on image resolution and telescope-specific characteristics. Our caption evaluations use only Legacy Survey images, and VLM descriptions could degrade for lower-resolution data where morphological features are less visually distinct or might push the model to make guesses about the unresolved morphology. Systematically benchmarking caption accuracy as a function of resolution and imaging modality is an important direction for extending this approach to future surveys.

We demonstrate that synthetic VLM descriptions provide a sufficient signal for semantic retrieval, significantly improving on image-similarity baselines and enabling open-vocabulary search. By training cross-modal encoders on $\sim$250,000 synthetic captions, we enable natural language search for over 100 million astronomical images without human annotation. A VLM-based re-ranking step further improves results, and performance increases with larger models and additional sampling. We establish that large-scale scientific image archives can be made semantically searchable without human supervision, offering a generalizable framework for scientific image retrieval that could extend to other domains with large unlabeled image archives, such as Earth observation, microscopy, and materials science. As VLM capabilities continue to improve, AI-generated annotations will become increasingly accurate and useful for scientific discovery.

\section{Code and Data Availability}
The code, trained model weights, datasets, and search application are publicly released to facilitate reproducibility and future research. The AION-Search model checkpoint, training data (galaxy descriptions and their text embeddings for Legacy Survey DR10 and HSC PDR3 Wide), pre-computed embeddings for over 100 million Legacy Survey galaxies, the VLM description benchmark, retrieval evaluation datasets (GZ-DECaLS embeddings with relevance labels and $k$-fold splits, lens retrieval embeddings, Galaxy10), and the catalog of stream candidates are available as a Hugging Face collection at \url{https://huggingface.co/collections/astronolan/aion-search}. The search application is hosted on Hugging Face Spaces for public use. The source code is available at \url{https://github.com/NolanKoblischke/AION-Search} and is archived on Zenodo~\citep{koblischke_2026_20438386}.
\facilities{DESI Legacy Imaging Surveys~\citep{legacy}, Hyper Suprime-Cam Subaru Strategic Program~\citep{hsc}}
\software{PyTorch~\citep{Ansel_PyTorch_2_Faster_2024}, NumPy~\citep{2020NumPy-Array}, pandas~\citep{The_pandas_development_team_pandas-dev_pandas_Pandas}, Matplotlib~\citep{Hunter_Matplotlib_A_2D_2007}, Astropy~\citep{Astropy_Collaboration_and_Price-Whelan_The_Astropy_Project_2022}, healpy~\citep{Zonca2019, 2005ApJ...622..759G}, h5py~\citep{andrew_collette_2023_7560547}, UMAP~\citep{mcinnes2020umap}}
\begin{acknowledgments}

We thank the anonymous referee for constructive comments that improved the quality and clarity of the manuscript. We acknowledge the Center for Computational Astrophysics at the Flatiron Institute for hospitality while a portion of this work was carried out. This work was supported by the AI2050 program at Schmidt Sciences (Grant G-25-70028) and by the Data Sciences Institute at the University of Toronto through grant number DSI-DSFY4R1P01. This research was enabled in part by support provided by Compute Ontario (\url{https://www.computeontario.ca/}) and the Digital Research Alliance of Canada (\url{https://alliancecan.ca}). Polymathic AI and SH gratefully acknowledge the support provided by Schmidt Sciences and Simons Foundation. We also thank Sebastian Wagner-Carena, Miles Cranmer, Francesco Pio Ramunno, Siddharth Mishra-Sharma, Tom Hehir, Tanya Marwah, Jacopo Teneggi, Mike Smith, Rafael Martínez-Galarza, Daniel Muthukrishna and Mike Walmsley for valuable discussions and helpful feedback that improved this work and Sophie Barstein for her help with writing the corresponding blog post.

This work makes use of data hosted on equipment supported by the Scientific Computing Core at the Flatiron Institute, a division of the Simons Foundation. It uses data from the DESI Legacy Imaging Surveys, including data obtained at the Blanco telescope (Cerro Tololo Inter-American Observatory, NSF's NOIRLab), the Bok telescope (Steward Observatory, University of Arizona), and the Mayall telescope (Kitt Peak National Observatory, NOIRLab), with support from NOIRLab, the Lawrence Berkeley National Laboratory, the U.S. Department of Energy, and the U.S. National Science Foundation. The complete acknowledgments can be found at \url{https://www.legacysurvey.org/acknowledgment/}.

This work makes use of data from the Hyper Suprime-Cam (HSC) Subaru Strategic Program, collected at the Subaru Telescope and developed by NAOJ, Kavli IPMU, the University of Tokyo, KEK, ASIAA, and Princeton University. The complete acknowledgments can be found at \url{https://hsc-release.mtk.nao.ac.jp/doc/index.php/acknowledging-hsc__pdr3/}.

Large language models were used to assist with editing and formatting this manuscript.
\end{acknowledgments}

\appendix
\section{Retrieval benchmark details}\label{sec:retrieval_data}
Retrieval experiments in \S\ref{sec:evaluating_search} use Legacy Survey images~\citep{legacy} following the AION retrieval protocol in Section 7.3 of \citet{2025aion1}, restated here for comparability. For spirals and mergers, we use Galaxy Zoo DECaLS, a citizen-science project where volunteers label galaxy morphology~\citep{2022walmsley}. We keep galaxies with at least three volunteer votes and cross-match with Legacy Survey South resulting in $\sim$170,000 galaxies. For evaluation, each candidate image gets a relevance score equal to the fraction of volunteers who chose that class ($r \in [0,1]$). We reproduce the AION parent sample for strong gravitational lenses by first cross-matching the Legacy Survey and HSC datasets to approximately reproduce the sample used in the HSC strong lensing searches~\citep{Jaelani2024}. Then we require: $0.2 \le z_{\mathrm{phot}} \le 1.2$ (photometric redshift), $M_\star > 5\times10^{10}\,M_\odot$ (stellar mass), and star formation rate per unit stellar mass less than $10^{-10}\,\mathrm{yr}^{-1}$. We then cross-match to published strong-lens catalogs as listed in \citet{2025aion1} Section 7.3 and assign relevance $r=1.0$ to cataloged lenses and $r=0.0$ otherwise, yielding 770 lenses.

Since lenses are extremely rare in this dataset, nDCG@10 is inherently noisy. The similarity-based baselines in Table~\ref{tab:spiral_merger_lens} average over many image queries per task ($\sim$29,000 for spirals, $\sim$690 for mergers, $\sim$770 for lenses), making their scores robust. AION-Search uses a single text query per category, so we cannot vary queries in the same way. To estimate variability, we randomly partition each evaluation set into 10 disjoint folds (without replacement) and recompute nDCG@10 on each fold for both AION-Search and AION-1-B. Table~\ref{tab:kfold_results} reports the mean and standard deviation across folds. AION-1-B scores are stable ($0.591 \pm 0.006$ for spirals, $0.226 \pm 0.029$ for mergers, $0.008 \pm 0.004$ for lenses), as expected from averaging over many queries. AION-Search achieves $0.973 \pm 0.033$ for spirals and $0.456 \pm 0.077$ for mergers, close to the full-dataset values. For lenses, the mean drops to $0.058 \pm 0.071$, below the full-dataset score of $0.173$: the full evaluation set places two confirmed lenses in the top 10, while folds average only $\sim$0.5 lenses in the top 10. Despite this variance, AION-Search remains approximately seven times higher than AION-1-B for lens retrieval ($0.058$ vs.\ $0.008$). To remain comparable with the original AION retrieval benchmark~\citep{2025aion1}, Table~\ref{tab:spiral_merger_lens} reports nDCG@10 values computed on the full evaluation set for all methods.

\begin{table}[h]
\centering
\begin{tabular}{lccc}
\toprule
& Spirals & Mergers & Lenses \\
\midrule
AION-1-B & $0.591 \pm 0.006$ & $0.226 \pm 0.029$ & $0.008 \pm 0.004$ \\
AION-Search & $0.973 \pm 0.033$ & $0.456 \pm 0.077$ & $0.058 \pm 0.071$ \\
\bottomrule
\end{tabular}
\caption{AION-Search consistently outperforms similarity search across 10-fold cross-validation. Mean nDCG@10 $\pm$ standard deviation across 10 random disjoint partitions of the evaluation set. AION-1-B averages over many image queries per category, yielding stable scores. AION-Search uses a single text query per category (``visible spiral arms'', ``merging'', ``gravitational lens''), so lens retrieval exhibits high variance due to the rarity of positives. Despite this variance, AION-Search remains substantially higher than AION-1-B across all categories.}
\label{tab:kfold_results}
\end{table}
\subsection{Lens retrieval sensitivity analysis}\label{sec:lens_sensitivity}
Some retrieved candidates that were not counted as lenses under the initial evaluation protocol in \S\ref{sec:evaluating_search} appear in other lens catalogs. The lens nDCG@10 scores in Table~\ref{tab:spiral_merger_lens} depend on two evaluation choices inherited from the AION retrieval protocol: which lens catalogs define the positive set, and the angular radius used to cross-match retrieved objects to catalog positions. The AION protocol uses HSC strong-lens catalogs with a 1$''$ match radius, focusing on the central object in each image. Because AION-Search is trained on VLM descriptions of the full image rather than just the central object, a wider match radius may be more appropriate. We also test adding \texttt{lenscat}~\citep{lenscat}, a community-contributed catalog containing additional lens candidates beyond those in the AION baseline catalogs.

Table~\ref{tab:lens_sensitivity} reports lens nDCG@10 under four configurations varying the catalog set (with and without \texttt{lenscat}) and match radius (1$''$ and 30$''$, approximately the image size). Both changes increase the number of matched lens candidates in the retrieved results and consequently raise nDCG@10 for both models. The effect is most pronounced when both are combined: on the full evaluation set, AION-1-B increases from 0.012 to 0.025 and AION-Search from 0.173 to 0.258. The 10-fold results show the same pattern. Because these evaluation choices affect both models comparably, AION-Search remains substantially higher than AION-1-B under all configurations, and our conclusions are not sensitive to these choices. Table~\ref{tab:spiral_merger_lens} reports the baseline configuration (HSC strong-lens catalogs, 1$''$) to remain comparable with~\citet{2025aion1}.
\begin{table}[h]
\centering
\begin{tabular}{llcc}
\toprule
Catalogs & Radius & AION-1-B & AION-Search \\
\midrule
\multicolumn{4}{l}{\textit{Full evaluation set}} \\
HSC strong-lens catalogs & 1$''$ & 0.012 & 0.173 \\
HSC strong-lens catalogs & 30$''$ & 0.019 & 0.173 \\
HSC strong-lens catalogs + \texttt{lenscat} & 1$''$ & 0.016 & 0.173 \\
HSC strong-lens catalogs + \texttt{lenscat} & 30$''$ & 0.025 & 0.258 \\
\midrule
\multicolumn{4}{l}{\textit{10-fold cross-validation (mean $\pm$ std)}} \\
HSC strong-lens catalogs & 1$''$ & $0.008 \pm 0.004$ & $0.058 \pm 0.071$ \\
HSC strong-lens catalogs & 30$''$ & $0.012 \pm 0.005$ & $0.080 \pm 0.079$ \\
HSC strong-lens catalogs + \texttt{lenscat} & 1$''$ & $0.012 \pm 0.006$ & $0.075 \pm 0.116$ \\
HSC strong-lens catalogs + \texttt{lenscat} & 30$''$ & $0.020 \pm 0.007$ & $0.173 \pm 0.144$ \\
\bottomrule
\end{tabular}
\caption{Relative advantage of AION-Search over similarity search is robust to evaluation protocol choices. Lens nDCG@10 under four configurations varying the positive lens catalog (HSC strong-lens catalogs alone or with \texttt{lenscat}; \citealt{lenscat}) and cross-match radius (1$''$, matching the central object, or 30$''$, approximately the image size). Both changes increase absolute scores for both methods, but AION-Search remains substantially higher than AION-1-B in all configurations. Table~\ref{tab:spiral_merger_lens} reports the baseline configuration (HSC, 1$''$) for comparability with~\citet{2025aion1}.}
\label{tab:lens_sensitivity}
\end{table}

\section{Prompts}
We provide the prompts used for caption summarization and VLM re-ranking in Figure~\ref{fig:summary_prompt} and Figure~\ref{fig:rerank_prompt}.
\begin{figure*}[h]
  \centering
  \begin{tcolorbox}[colback=gray!10, colframe=black, title=Summarizer prompt]
Please summarize the following description into a single sentence CLIP query:
\begin{verbatim}
<original_description>
{{original_description}}
</original_description>
\end{verbatim}
Only output the summary without any additional text.
Do not use any of the following words: `no`, `not`, `without`, `absence`, `lack of`, `no obvious`, `no signs`, `absence of` or any other negation words or phrases. 
Ignore any phrases containing negation from the original description containing these words.
\end{tcolorbox}
\caption{Prompt for generating single-sentence summaries from multi-paragraph VLM descriptions using GPT-4.1-nano. The prompt leads to shorter descriptions that exclude negations to improve contrastive learning performance~\citep{li2023an}.}
\label{fig:summary_prompt}
\end{figure*}
\begin{figure*}[h]
  \centering
  \begin{tcolorbox}[colback=gray!10, colframe=black, title=Re-ranking prompt]
Does this galaxy image display signs of gravitational lensing? Rank 1-10 where 10 means you are entirely sure there are signs of gravitational lensing and 1 being you are entirely sure there are no signs of gravitational lensing.
\end{tcolorbox}
\caption{Prompt used in the VLM re-ranking stage for the experiment in \S\ref{sec:reranking}: for each retrieved galaxy image, the model assigns a 1–10 score indicating confidence in gravitational lensing.}
\label{fig:rerank_prompt}
\end{figure*}
\bibliographystyle{aasjournalv7}
\bibliography{sample701}{}
\end{document}